# A single platform with van der Pauw geometry for measurement of Seebeck coefficient, resistivity, and Hall effect of thin films

Niraj Kumar Singh[1]*, Martin Falk[2,3], Per-Anton Schön Wallander[2], Per Sandström[2], Per Eklund[1,2] and Arnaud le Febvrier[1]

*[1]Department of Chemistry – Ångström, Uppsala University, SE-751 21, Uppsala, Sweden*
*[2]Thin Film Physics Division, Department of Physics, Chemistry, and Biology (IFM), Linköping University, SE-581 83, Linköping, Sweden*
*[3]Present address: Impact Coatings AB, Cobolgatan 5, SE-583 30 Linköping, Sweden*

**Author to whom correspondence should be addressed: niraj-kumar.singh@kemi.uu.se*

## Abstract:

A modular thermoelectric properties measurement setup in van der Pauw configuration was developed for a straightforward and simultaneous measurement of electrical resistivity and Seebeck coefficient in an extensive temperature range of 25°C - 600°C and can also perform Hall measurements at room temperature. The setup is optimized for accurate measurement of voltages and temperatures gradients by minimizing possible errors from offset voltages, wire contributions and thermal contact resistances which helps getting reliable data. The setup is user friendly, and the measurements are fully automated and controlled using a LabVIEW program. The detachable modules make this setup quite versatile and provide an all-in-one (except thermal conductivity) solution for thermoelectric measurements.

# Introduction

The ever-rising global energy demand and limited conventional sources of energy has attracted immense interest for the search of alternate sources of energy as well as scavenging waste heat. [1] Thermoelectric materials can directly convert heat energy into electricity and are seen as a green and alternate source of energy. The conversion efficiency of a thermoelectric device is gauged by a dimensionless parameter known as figure of merit, $ZT = \frac{S^2\sigma}{k}T$, where $S$ is the Seebeck coefficient, $\sigma$ is the electrical conductivity, $k$ is the total thermal conductivity and $T$ is absolute temperature.[2-4] Most of the research in thermoelectrics revolves around optimizing and consequently achieving a fine balance between these adversely related transport properties. Apart from thermoelectrics, Seebeck coefficient ($S$) and electrical transport properties such as electrical conductivity ($\sigma$), charge carrier concentration ($n$), and carrier mobilities ($\mu$) are widely studied for various other studies and applications.

For the optimization of ZT, Seebeck coefficient $S$ is an important factor and challenging to quantify accurately. Seebeck coefficient, S ($-\frac{\Delta V}{\Delta T}$) is defined as the potential, ΔV developed across two points due to a temperature gradient, $\Delta T = T_2 - T_1$, where $T_2$ and $T_1$ is the temperature of the hot side and cold side, respectively.[5] The measurement of S appears quite straightforward since it only requires voltage and temperature measurements.[6] However, in practice it is a challenging task due to the intricacies involved during the measurement. Generally, the measurement is done in a two-probe configuration and the presence of residual thermoelectric voltages (μV) adds up to the inaccuracies.[7] Measuring thermoelectric transport properties of thin films using conventional commercial instruments, which are generally designed for bulk samples, is challenging due to the anisotropic nature of the samples, the size, and are usually time consuming. Hence, the development of a versatile setup for measurement of thermoelectric transport properties ($S$, $\sigma$, $n$ and μ) in an extended range of temperature is necessary.

There are two primary methods for the measurement of Seebeck coefficient, namely: (a) integral method and (b) the differential method. In the case of integral method, one end of the sample is continuously heated without changing the temperature of the other end. [8, 9] Differentiating the ΔV with respect to the hot side temperature gives S. This allows the achievement of a larger ΔT across the sample. However, this technique is not suitable for smaller samples due to the onset of thermal back diffusion which restricts them from getting higher ΔT. For smaller samples, differential method is used in which a chosen ΔT is maintained across the two ends of the samples by simultaneously increasing the temperatures of both the ends and measuring the ΔV which gives $S = -\frac{\Delta V}{\Delta T}$.[10, 11] The differential method for Seebeck measurement can further be classified as (a) steady-state,[12] (b) quasi-steady-state[13] and (c) transient method (AC)[14, 15]. Under the steady-state method, a stabilized ΔT is achieved by maintaining desired hot ($T_h$) and cold ($T_c$) side temperatures and ΔV is recorded across the sample. This process often takes much longer time and hence it is usually not preferred. To reduce the measurement durations, quasi-steady-state approach can be adopted. In that case, a gradient heater provides an increasing heat flux while continuously measuring the ΔV across the sample. Proper implementation of gradient heating is a tedious task and requires several high impedance voltmeters which makes the setup complicated and costly. On the other hand, the transient-ac method is a technique more suitable for high resistivity samples. Due to these reasons, steady-state method is widely used for instrumentations including most of the commercial systems.[16-19] Most of the commercial setups provide reliable measurement of data but they also suffer few drawbacks such as limitation of sample size, relevant temperature range of measurement and/or small enough step size of data recording. Therefore, designing a reliable and customized TE properties measurement system for thin films is very important.

There already exists number of articles on designing TE properties measurement system dedicated for thin films samples.[7, 20-22] Dörling *et al.*, designed an all-in-one setup for measuring $S$ and $\sigma$ of anisotropic thin films based on van der Pauw method.[23] Pereira *et al.* developed a customized setup for the measurement of $S$ and $\sigma$ for thin films in the cryogenic temperature range of 10-300 K with a high accuracy (within ~ 3%) for the studied sample.[24] Another study reports on setup for measuring $S$ and $\sigma$ which was designed to work on both thin films as well as bulk samples, in the temperature range of 77 – 330 K.[25] In addition to measurement setups designed to work below room temperature, there are reports on setups which can measure to higher temperatures. A measurement setup reported by Sarath *et al.*, using the integral method is capable of accurately measuring $S$ of thin films in the temperature range of 300 – 650 K.[9] Overall, we have a rich literature on designing customized thermoelectric measurement setups, however, they are limited in measurement capabilities such as working range of temperature or lacking a single platform for measuring $S$, $\sigma$, $n$ and μ.

The setup described in this article is straightforward, inexpensive, and versatile. It can measure in-plane electrical resistivity, Seebeck coefficient and can perform Hall measurements on a square sample (typical size of 10 × 10 mm$^2$) in van der Pauw configuration in an extensive temperature range of 25°C - 600°C. This setup is suitable for measuring thin films grown on insulating substrates, which produces reliable data by minimizing any chemical interaction between the film and the substrate. The setup can reliably characterize thin films down to ~ 60 nm. For films thinner than this, the chance of damage is higher while handling the contacts. The base setup provides versatility for attaching different modules for either simultaneous measurement of Seebeck coefficient and resistivity or Hall analysis. The detachable modules make this setup quite versatile and provide an all-in-one (except thermal conductivity) solution for thermoelectric measurements. The design is also capable of performing measurements in vacuum or inert environment making it suitable for oxygen-sensitive samples.

# Method

Methods in this article consist of two measurements: (a) simultaneous measurement of electrical resistivity and Seebeck coefficient, and (b) Hall measurement. Measurement of resistivity is based on the well-established van der Pauw method, whereas Seebeck coefficient is measured using the steady-state method. The Hall effect measurements are performed at room temperature with a fixed magnetic field.

## Electrical resistivity measurement

In its conventional form, the van der Pauw method is used to determine the sheet resistance $R_S$ of the film, from which the electrical resistivity $\rho$ can be obtained if the sample thickness $t_S$ is known:

$$\rho = t_S \times R_S \tag{1}$$

Four electrical contacts are made at the periphery of the sample, here numbered (1-4). The resistivity measurement is performed by driving current through two electrodes while the voltage drop is measured across the other two electrodes. To eliminate any artefacts of measurement, such as thermoelectric voltages and magnetoresistance contributions, voltages are measured in eight different configurations as shown in figure 1.

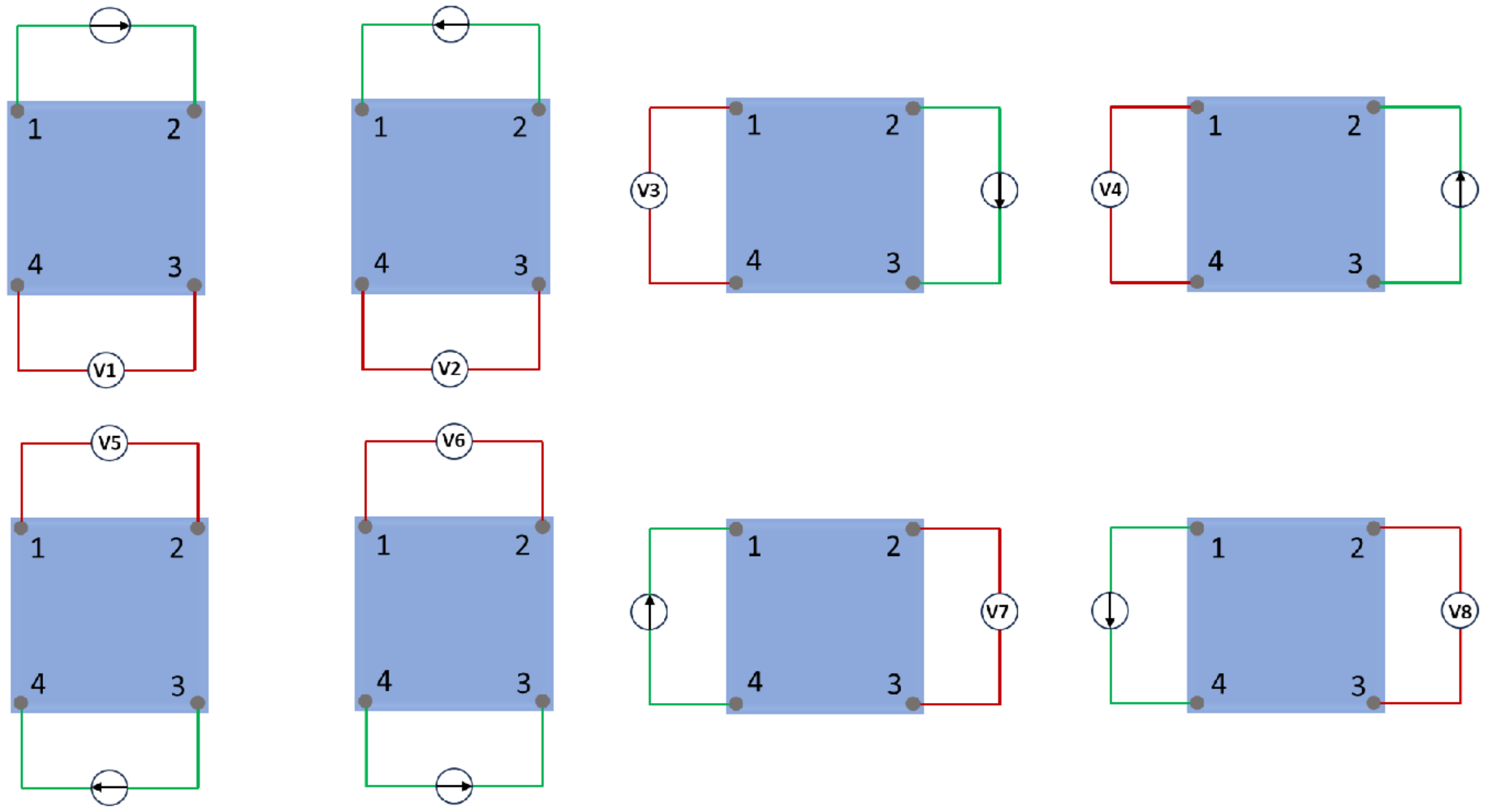

**Figure 1:** Geometries of the sample showing the position and the numbering of the four electrical contacts in the configuration for measuring the electrical resistivity measurement.

The eight measured voltages are then used to calculate the sheet resistance $R_S$ based on the following equations:

$$R_S = \frac{\pi}{ln2} \frac{(R_A + R_B)}{2} f(r) \quad (2)$$

$$R_A = \frac{(V_{34,12} - V_{34,21} + V_{12,34} - V_{12,43})}{4\ I} \quad (3)$$

$$R_B = \frac{(V_{41,23} - V_{41,32} + V_{23,41} - V_{23,14})}{4\ I} \quad (4)$$

where $R_A$ and $R_B$ represent the average resistance (Ω) measured in the two orthogonal directions. The factor $\frac{\pi}{ln2}$, appears after simplification of the van der Pauw's equation. $V_{34,12}$ represent the voltage drop (V) measured between probe 4 and 3 while the current is injected across 1 and 2, etc.; $I$ represents the injected current (A) which is the same for all eight voltage measurements; and $f(r)$ is the geometrical correction factor based on sample geometry which is related to the resistance ratio $R_A/R_B$. For an isotropic square shaped sample, we have $R_A = R_B$ which gives $f(r) = 1$.[26, 27]

## Hall coefficient measurement

Understanding thermoelectric materials requires information about the nature of charge conduction and the ability to tune carrier concentration. The carrier concentration, conduction type (*n* or *p* type, or mixed), and the carrier mobility can be derived from the measurement of Hall coefficients, performed by applying a magnetic field perpendicular to the surface of measurement.

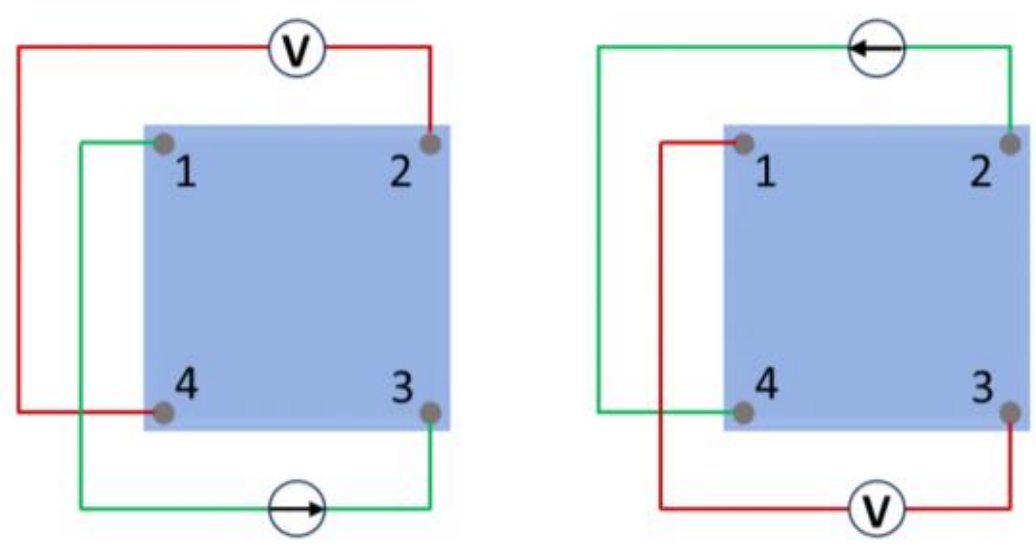

**Figure 2:** The geometry of the sample showing the position and numbering of electrodes for Hall coefficient measurement.

The Hall coefficient measurement is carried out in a similar fashion to the classic van der Pauw resistivity measurement. However, the current is injected diagonally into the sample while the voltage drop is measured in the other diagonal contacts (Figure 2). The voltage drop is measured in a similar fashion to the van der Pauw method for minimizing residual voltages. Eight voltages are measured: (i) when the magnetic field has a positive polarity ($B+$), two voltages ($V_{24,13}^{B+}$ and $V_{31,24}^{B+}$) plus their corresponding voltages with reversed current ($V_{42,31}^{B+}$ and $V_{13,42}^{B+}$); and similarly, four more voltages are measured when the magnetic field has a negative polarity ($B-$). Using the eight measured voltages, the average Hall coefficient ($R_{Havg}$) in $cm^3C^{-1}$ can be calculated using the following equations:

$$\sum V_{Hall} = V_C + V_D + V_E + V_F \tag{5}$$

$$V_C = V_{42,13}^{B+} - V_{42,13}^{B-} \tag{6}$$

$$V_D = V_{24,31}^{B+} - V_{24,31}^{B-} \tag{7}$$

$$V_E = V_{31,24}^{B+} - V_{31,24}^{B-} \tag{8}$$

$$V_F = V_{13,42}^{B+} - V_{13,42}^{B-} \tag{9}$$

$$R_{Havg} = \frac{\sum V_{Hall}}{8} \times \frac{t_s}{I\,B} \quad (10)$$

where $V_{42,13}^{B+}$ represent the voltage drop (V) measured between probe 4 and 2 while the current is injected across 1 and 3 with a positive magnetic field, etc.; $I$ represent the injected current (A); $t_s$ represent the thickness of the sample (in cm); $B$ is the magnetic flux density. Usually, B is given in SI unit T, while distances are given in cm rather than $m^2$, thus the formula above needs to be multiplied by $1.10^4$ for the conversion from $m^2$ to $cm^2$. The average Hall coefficient is then used to calculate the charge carrier concentration and the Hall mobility using the following formula:

$$n = \frac{1}{q\,|R_{Havg}|} \quad (11)$$

$$\mu_{Hall} = \frac{|R_{Havg}|}{\rho} = \frac{1}{\rho\, n\, q} \quad (12)$$

where n is the carrier concentration (in $cm^{-3}$; positive and negative $R_{Havg}$ represents holes and electrons as majority charge carriers, respectively), q is the elementary charge, $\mu_{Hall}$ is the Hall mobility (in $cm^2V^{-1}s^{-1}$) and $\rho$ is the resistivity in (Ω.cm).

## Seebeck coefficient measurement

Measurement of *S* requires electrical as well as thermal measurements, a challenge because of the need for accurate and precise measurement of the temperature. Moreover, the observation of non-zero ΔV along the contacts, at ΔT = 0, is the most common error while measuring S. Additional sources of error can be parasitic heat flux due to the thermocouples, and contact resistances.[25]

In the present setup, $S = \frac{\Delta V}{\Delta T}$ measurement is performed by (i) applying a temperature gradient (ΔT) along the diagonal of the samples (10 x10 $mm^2$) using a heater placed at one of the corners at the backside of the substrate, and (ii) measuring the generated voltages and temperatures through the same diagonal. The temperatures and the voltages ($V_{total}$) are measured at the same position because the voltages are measured between the Chromel® legs of both thermocouples. The total measured voltage is the sum of the thermoelectric voltage from the sample and Chromel® wires. The thermoelectric voltage from the sample and its Seebeck coefficient can be extracted as follows:

$$V_S = \Delta V_{total} + S_{ch}.\Delta T \quad \text{or} \quad S_S = \frac{\Delta V_{total}}{\Delta T} + S_{ch} \tag{13}$$

where $S_{Ch} = 21.035 + 2.396 \times 10^{-2}T - 6.44 \times 10^{-5}T^2$.[28] For achieving higher measurement precisions, Seebeck coefficient is measured by averaging three values (S1, S2 and S3) extracted from the measured voltages at three different temperature gradients ($\Delta T_1$, $\Delta T_2$ and $\Delta T_3$).

## Design and measurement setup

The entire measurement setup can be divided into three parts (Figure 3): (i) the vacuum chamber, (ii) the measurement module, and (iii) the measurement unit (SMU) including the computer-controlled data acquisition system with management of both the input data and measurement apparatus. The measurement modules are interchangeable modules allowing the measurement of the electrical resistivity of the sample up to 1000°C or the Seebeck coefficient and electrical resistivity up to 550°C. The two modules use the same base setup with a vacuum tube, which can be placed in a tube furnace, and use the same computer-controlled data acquisition and management of both the data input and measurement apparatus.

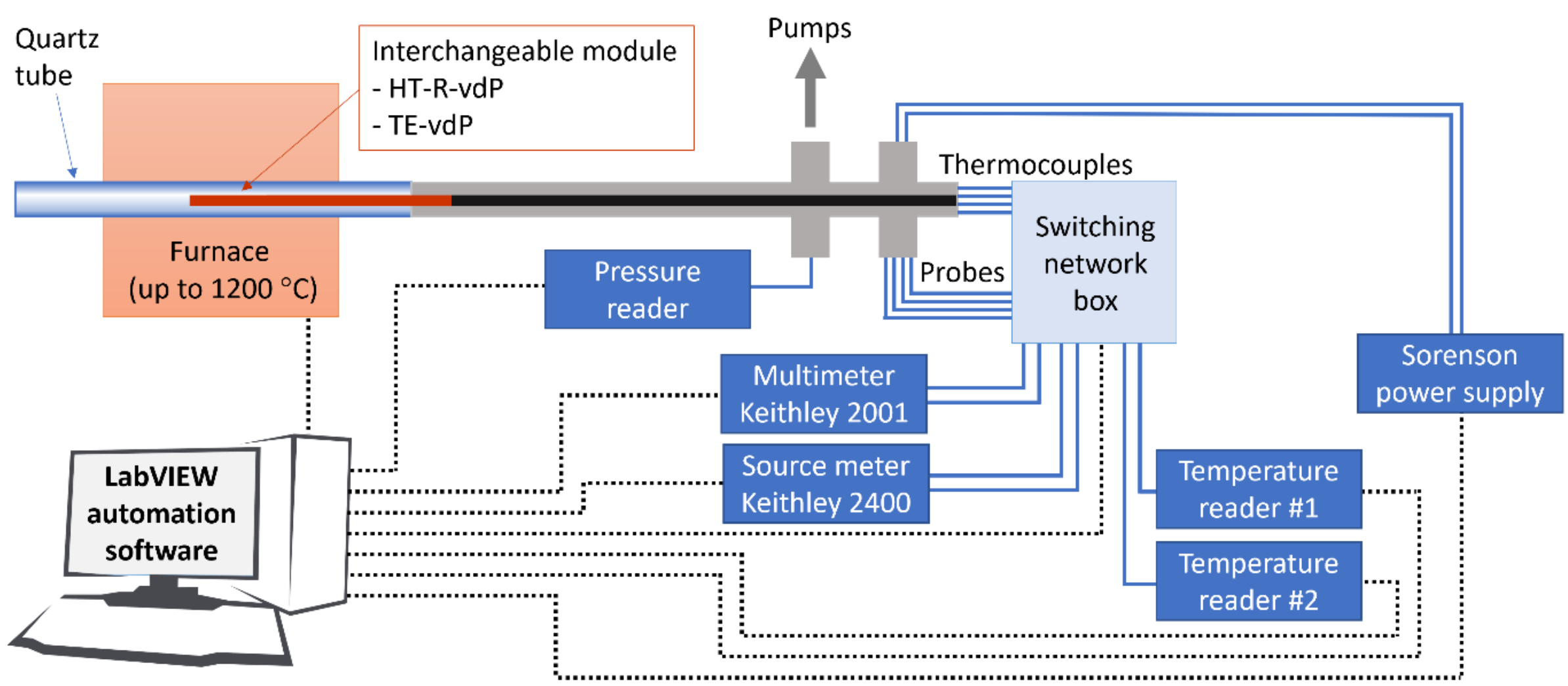

**Figure 3:** Schematic view of the overall measurement setup. The vacuum chamber/furnace (grey-orange) with integrable modules (red) and the SMU (blue) connected to a LabVIEW program for automation.

The vacuum system consists of a combination of dry Scroll pump (Pfeiffer HiScroll 6, pumping speed ~ 100 l/min) and a turbomolecular pump (Pfeiffer TPH 190, pumping speed ~ 190 l/sec).

The vacuum chamber, with a base pressure of $3\times10^{-3}$ Pa ($2\times10^{-5}$ Torr), consists of a long cylindrical tube of stainless steel 316 (∅ 45 mm × 45 mm) with a long quartz tube (∅ 35 mm). Only the quartz tube is placed in a tube furnace while the rest of the vacuum chamber stays far from the furnace and stays close to room temperature. The furnace (MTI OTF-1200X) is a split tube furnace which can reach a temperature as high as 1200 °C with a constant temperature zone of 100 mm in the center with an accuracy of ± 2°C at 900 °C. Most of the flanges for the thermocouples and electrical connections are as far as possible from the furnace.

## High Temperature Resistivity van der Pauw module (HT-R-vdP)

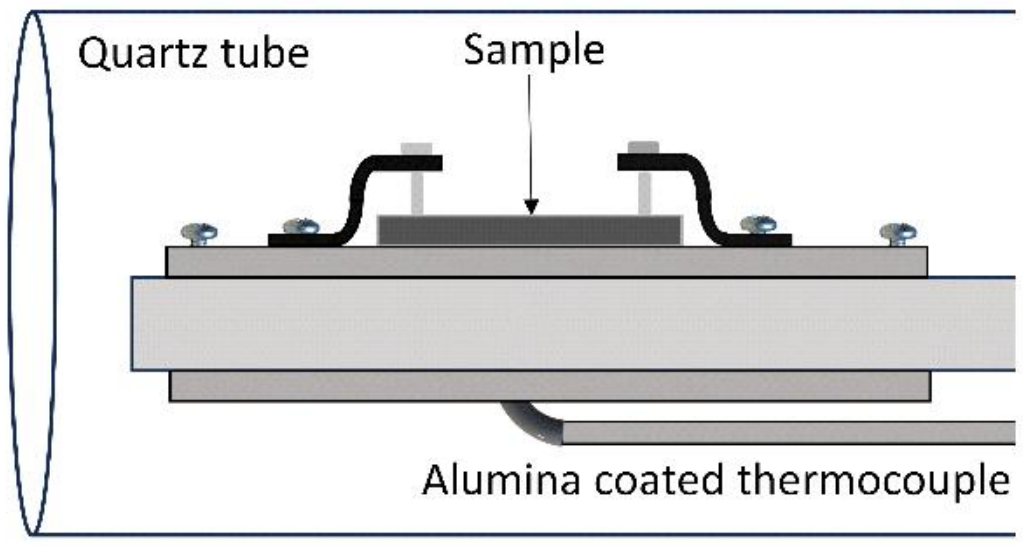

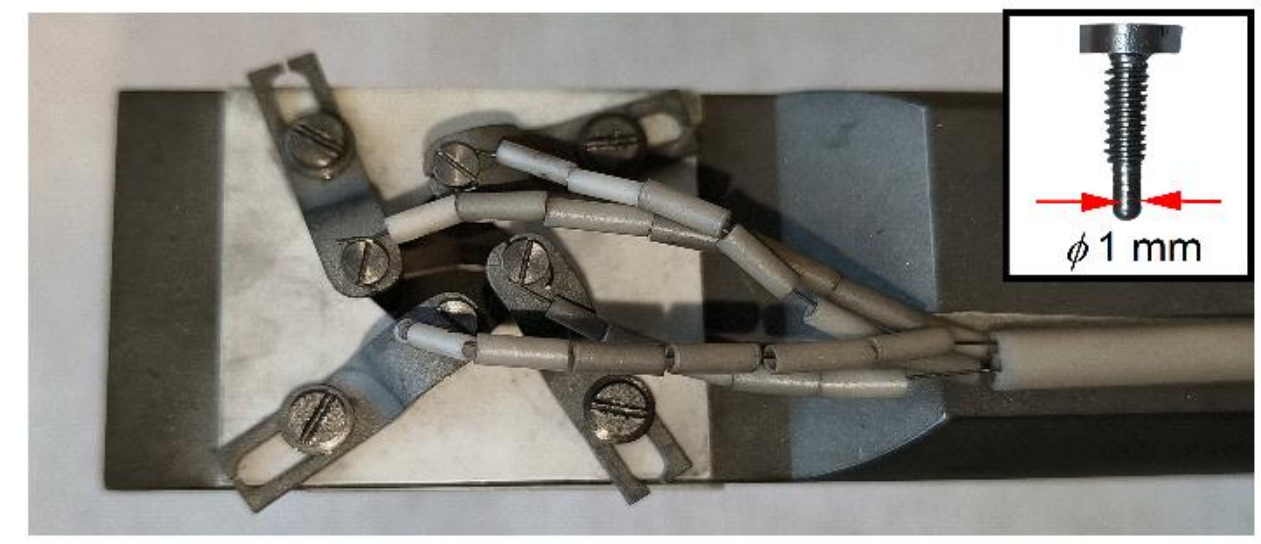

**Figure 4:** (a) Schematic representation, and (b) actual photograph, of the high temperature resistivity measurement module (HT-R-vdP). The inset of Fig. 4(b) shows that the probes 1 mm in diameter and has rounded tips.

Figure 4 shows a schematic view of the high temperature resistivity van der Pauw module. The assembly is entirely made of chemically stable tungsten components and high purity alumina plates, both of which are widely used for high-temperature measurements in vacuum and controlled atmospheres. Furthermore, the investigated samples are stable nitride thin films (e.g., CrN and ScN) that contain no volatile constituents, thereby eliminating the possibility of sample-induced contamination during measurements. Hence, the observed blackening of the setup is not due to contamination or degradation. While the discoloration is visually noticeable due to prolonged thermal cycling, it does not affect the electrical or thermal performance of the setup. The electrodes consist of rounded end screws (∅ 1mm, Inset of Fig. 4(b)) in tungsten, and they are mechanically connected to a (∅ 0.125-mm) tungsten wire. Thin alumina plates at top and bottom are used as well as several alumina washers to play the role of electrical insulator limiting any short circuit between the electrodes and the rest of the assembly. Note that the design of the electrode allows for sample size variation between $5\times5$ mm$^2$ to $12\times12$ mm$^2$. For mounting samples, the different electrodes are fixed by softly tightening the screws at the base to create a soft spring effect for contact. The temperature measurement is performed with a thermocouple paced at an equivalent symmetrical position as the sample. The HT-R-vdP module allows the measurement resistivity up to high temperature up to 600 °C in vacuum when the module is placed inside a tube furnace.

The exact same module can be used for the room temperature Hall measurement where the module can be slid in between two magnets (Figure 5). The magnets are mounted in a rotating aluminum frame with a gap of cross-section 35×60 $mm^2$ allowing the placement of the module at the center. The magnetic field at the sample position was measured to be providing a magnetic field of 0.485 T.

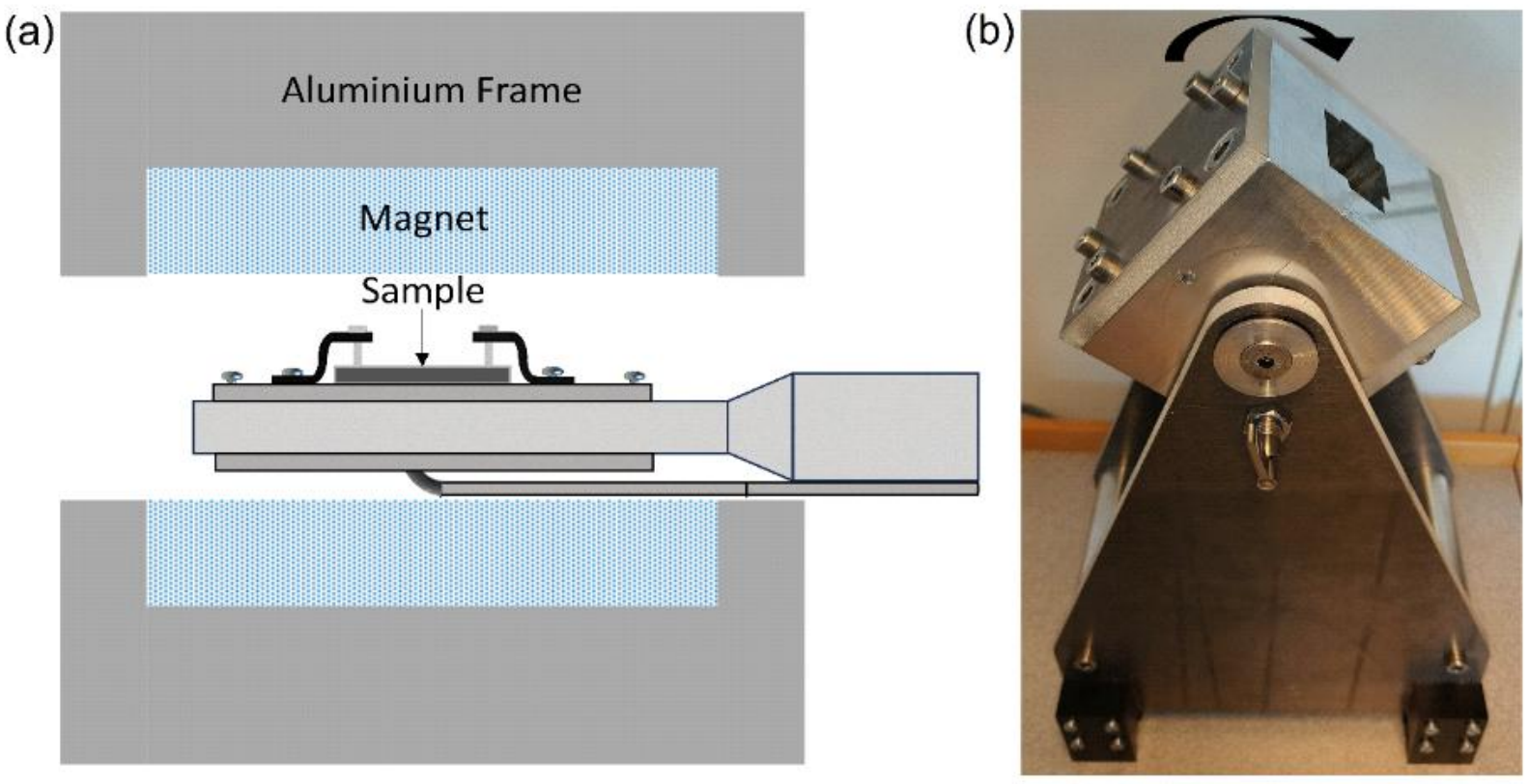

**Figure 5**: (a) Schematic representation of Hall measurement setup with the HT-R-vdP module and (b) magnet assembly with option to reverse the direction of magnetic field.

## <u>Thermoelectric van der Pauw module (TE-vdP)</u>

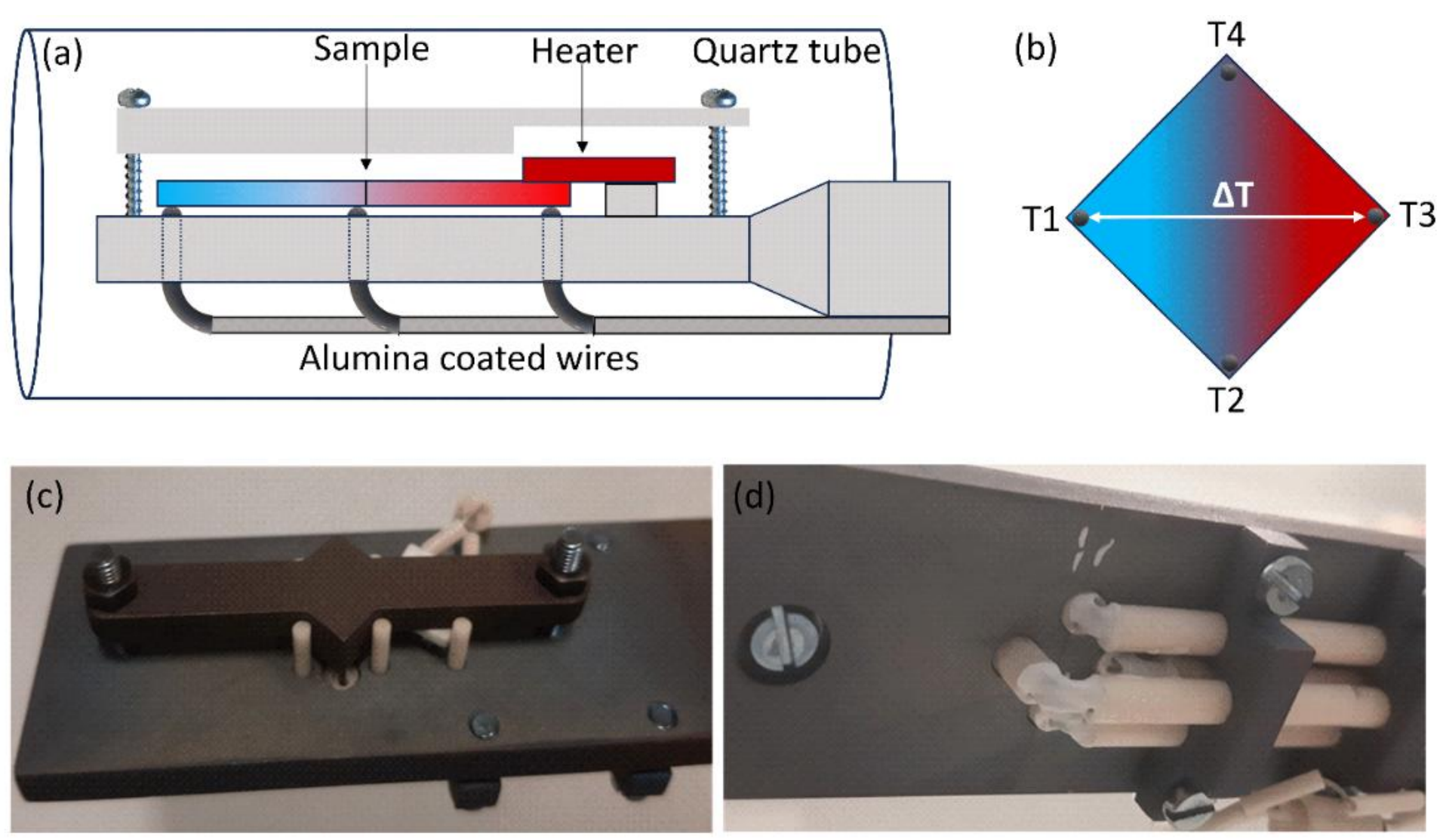

**Figure 6:** Schematic presentation of (a) side view of TE-vdP module with a sample mounted and (b) temperature gradient present on the sample with the electrode/thermocouple name. Actual images of the (c) top and (d) bottom view of the module.

Figure 5 shows a schematic view of the thermoelectric van der Pauw module (TE-vdP). The assembly is made of stainless steel SS-316, chosen for its stability up to 650 °C with low outgassing rate (~ $10^{-8}$ Pa-$m^3s^{-1}m^{-2}$) in vacuum.[29] The four electrodes consist of four K-type thermocouples (∅ 1 mm). Each thermocouple is electrically insulated, and any connector used for the thermocouple are in Chromel® or Alumel® to avoid the creation of residual Seebeck voltages with a material difference. The four electrodes are placed in fixed positions forming a square of 8x8 $mm^2$. The sample is mounted upside down with the film touching the thermocouple and it is fixed with a backing plate to ensure the connection. For creating a temperature gradient, a 10x10x1$mm^3$ resistive heater is placed behind the sample at one of the corners.

The temperature measurement is a crucial step for the Seebeck coefficient. To minimize the influence of external sources of heat (or heat loss), the thermocouple wires must be entirely thermally insulated from the point of measurement to the flange. Therefore, a high purity alumina paste was used to cover any naked thermocouple wire parts that can appear at ceramic junctions or bent areas.

## Computer-controlled data acquisition

The SMU consists of a Keithley 2400 Source meter, a Keithley 2001 multimeter, a Sorenson dc power supply, two temperature measurement units (Omega, DP41-TC) and a pressure gauge (Pfeiffer, TPG 361). The source meter (Keithley 2400) can deliver current from 1μA to 1 A (but limited to 40 mA in this setup) with a preprogramming resolution of 0.05%. The multimeter (Keitley 2001) allows the measurement at a microvolt precision with 7.5 digits resolution (± 0.5 μV) and the temperature reader has an accuracy of ± 0.005%.

In between the SMU and the chamber, a switching network box is used to be able to perform the different measurements required for van der Pauw or Seebeck measurement. The switching network box consists of series of relays allowing the automation for the selection of the probe paths for delivering the current and measuring the drop voltages. All the different parts (furnace, SMU, switching network box) are computer controlled through a LabVIEW program allowing high modularity and individual control over the various measurements performed with the apparatus. A more detailed description of the program can be found in supplementary information.

# Measurement procedures

## Temperature dependence resistivity using the HT-R-vdP module.

Prior to each measurement, I(V) tests are performed between each probe to confirm an Ohmic contact between the probes and the surface of the sample. The current is ramped up from 50 μA to 10 mA and the linearity between measured voltage and the delivered current is checked between two probes and repeated for each combination of probes. The temperature dependent electrical resistivity measurement is conducted in vacuum to avoid any possible reaction between the sample and the ambient air. To reach good vacuum ($< 4\times 10^{-5}$ torr / $5 \times 10^{-3}$ Pa), the pumping needs to be initiated prior to starting the measurement. Often a step of pre-degassing at 120-150 °C helps desorbing any molecules that could have absorbed on the surface or the sample stage.

The van der Pauw resistivity measurement is conducted at a fixed current level, which is determined by user. The current needs to be high enough to ensure a measurable voltage output by the Keithley 2001 multimeter. A higher current is preferable for highly conductive films, while a lower current will be more suitable for films with higher resistivity. The van der Pauw (a cycle of eight measurements) measurement is performed without interruption and is repeated constantly as many times as the user needs. Each cycle will generate one electrical resistivity data point every few seconds. The furnace temperature program is generated independently and can be adjusted as the user desires with a simple increase/decrease ramping, or more complex temperature profiles. This configuration allows a lot of freedom in terms of temperature profile measurement but one must be careful with the level of temperature ramping in comparison with the duration of one measurement cycle.

### Correction, optimization, and limitations

- Sample size correction: The van der Pauw method offers numerous advantages compared to other 4-point probe methods for electrical resistivity measurement. As explained in section II(a), with the appropriate measurement sequence, the thermoelectric and magnetoresistance residual voltages are rectified. The main correction in the van der Pauw measurement is the geometrical correction factor due to probe positioning or sample size. This correction is implemented automatically through the LabVIEW code.[27]

- Temperature limitation: Because the thermocouple is placed at a symmetrically equivalent position beneath the sample, the reading temperature and the sample temperature can differ during rapid temperature increase or decrease. Therefore, to minimize this error source, a maximum temperature ramping at 10 °C/min is recommended for the measurement of the temperature dependent resistivity.

- Electrical resistivity limitation: Theoretically, with (i) the combination of the Keithley 2001 and Keithley 2400 and (ii) the maximum current limit set at 40 mA due to the wiring, the limit of the measured sheet resistance is from 10 mΩ/□ to 100 MΩ/□.

## Hall measurement using the HT-R-vdP module

Prior to each measurement, and like the electrical resistivity measurement, a set of I-V tests are performed to confirm the ohmic character of the contacts. The HT-R-vdP module is placed in between two permanent magnets (∅ 50 mm each) separated by 35 mm which creates a magnetic field of 0.485 T at the sample position. First, using the automation LabVIEW software, the four voltages are measured with a positive magnetic field, then the magnets are flipped to deliver a negative magnetic field, and the four remaining voltages are measured.

## Correction, optimization, and limitations

- Sample positioning: During Hall effect measurements, errors may arise due to positioning of the sample in the magnetic field. This error is minimized with a guide tool which helps to contain the error of positioning of the sample holder within 1 mm in x, y and z directions. To estimate the error in $n$, a series of Hall measurements were performed by deliberately misplacing the sample along the x direction from its ideal position (Fig. 7). For a reference sample (CrN), a difference of ~20 % was observed when the sample was deliberately misplaced to ± 20 mm from its correct position, a position which would not be feasible by a user. Nevertheless, in the current apparatus, a normal human error of a maximum ± 1 mm from its ideal position is expected which yield an error of <1% for the reference CrN film.

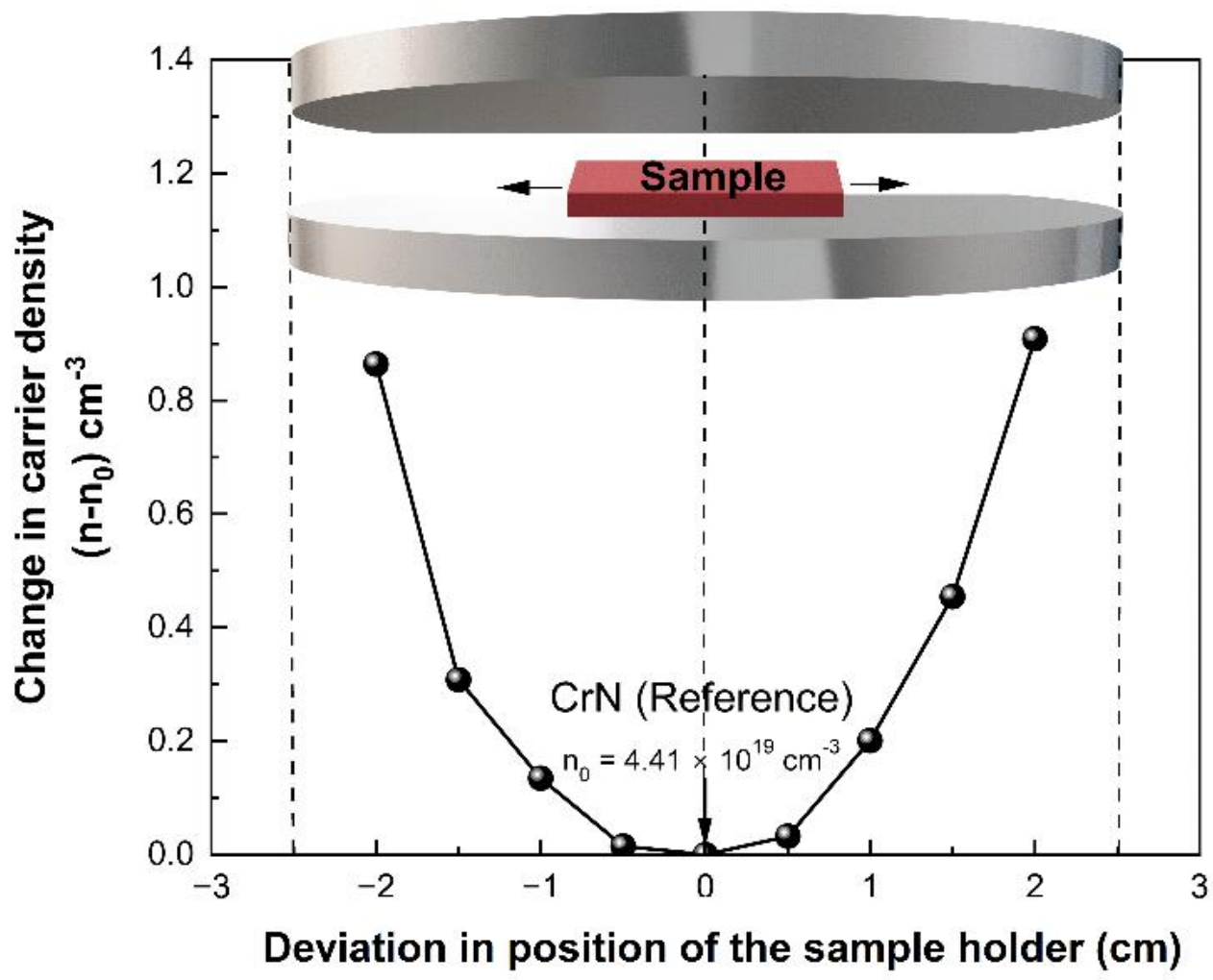

Figure 7: Change in charge carrier density, n, for s reference 220 nm CrN film grown on c-sapphire substrate, as a function of deviation in the position of sample holder from ideal central position during Hall measurements.

- Sample carrier concentration limitations: For the example of a 100 nm thick film, with the SMU and the magnetic field limitations, theoretically, the setup has a detection limit of carrier concentrations between of $1\times10^{18}$ cm$^{-3}$ and $1 \times 10^{24}$ cm$^{-3}$.

## Temperature dependence Seebeck and resistivity using the TE-vdP module.

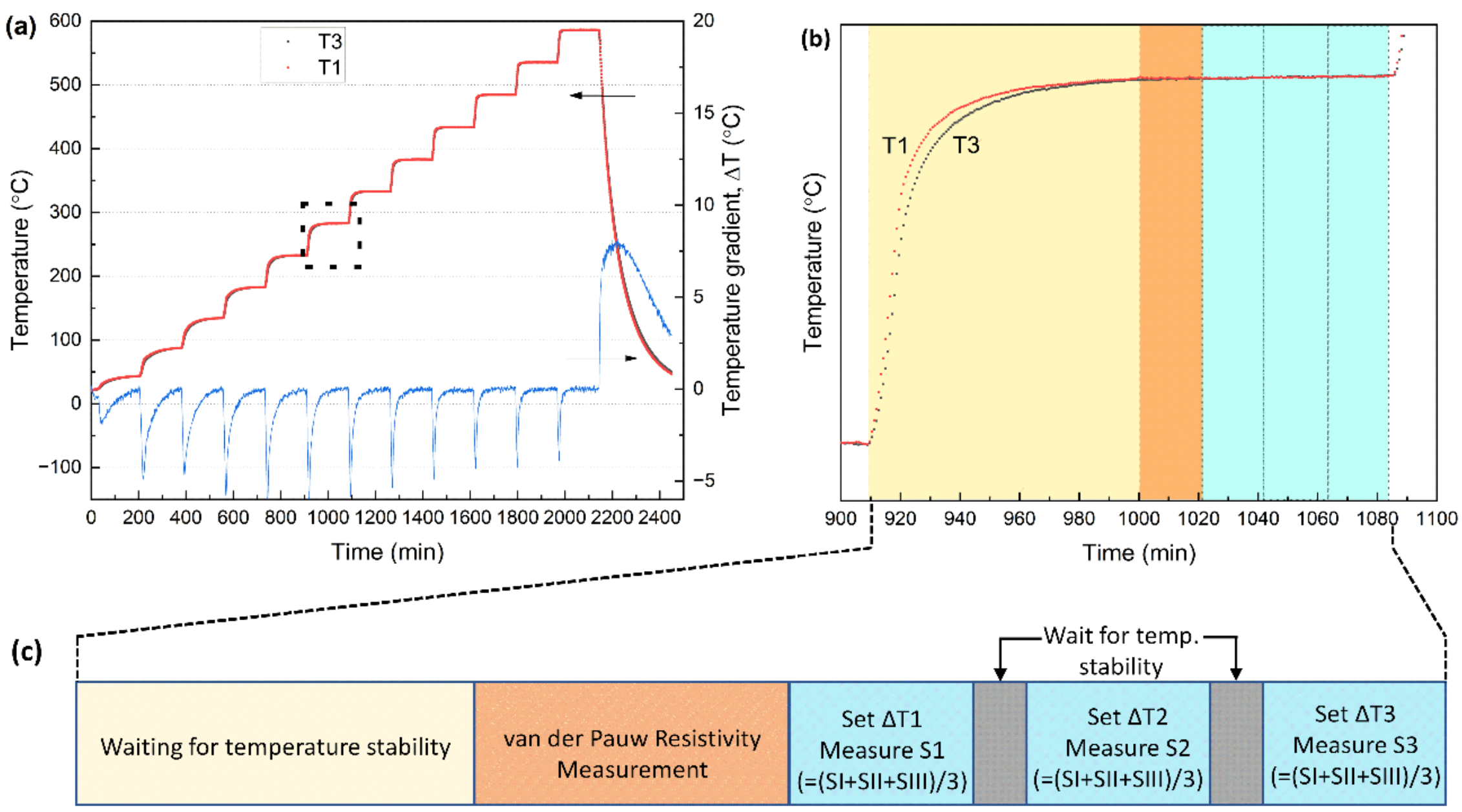

Figure 8: A temperature-time graph during a test Seebeck coefficient measurement showing (a) multiple steps during a complete measurement cycle from room temperature to 600 °C, with a step size of 50 °C, and (b) an enlarged view of a single step showcasing different procedures such as (c) waiting time for attaining temperature stability followed by resistivity measurement and Seebeck measurements.

Prior to TE measurement and like the electrical resistivity measurement, an I-V test is performed to confirm the Ohmic character of the contacts. Figure 8(a) shows a systematic temperature profile for a test TE measurement with multiple temperature dwelling and ramping steps. Further, after reaching temperature stability at every set temperature, ρ and S measurement takes place in a specific order as shown in Fig. 8(b-c). Once a stable set global temperature is obtained using the external furnace, the electrical resistivity is measured first using the van der Pauw method as explained in section II(a). The Chromel® wire from the K-type thermocouple is used to deliver (measure) the current (voltage). Afterward, a small heater mounted placed in the sample holder is activated to create the first temperature gradient (ΔT1) between the probes 3 and 1 (Figure 6) and followed by a waiting time allowing a stabilization of the temperature gradient. Once the temperature gradient is stable, a sequence of measurements is performed in the following order to record: (i) temperature at probes 1 and 3; (ii) voltage drop between the probes 1 and 3; (iii) both repeated for a total of three times. The sequence is then repeated for two other temperature gradients (ΔT2 and ΔT3). To eliminate possible errors during the measurement which may arise due to residual voltages, the recorded voltages are plotted as a function of ΔT, and the slope of the linear fit provides the corrected sample Seebeck ($\Delta V_{total} = S_{sample} \times \Delta T + \Delta V_{residual}$). Since the voltages are recorded for a total of three times at every $\Delta T$, three $\Delta V$ vs $\Delta T$ plots result in three Seebeck values, which are averaged to get the final S. This marks the end of measurement sequence at a particular global temperature, and the program ramps up the temperature for the next set value where the same set of measurements take place generating temperature dependent Seebeck data.

## Corrections, optimization, limitations

Several corrections need to be taken into consideration in Seebeck coefficient measurement. The main source of error of measurement comes from the fact that residual thermoelectric voltages alter the reading values. The temperature readings are the first affected but also the voltages because the Seebeck coefficient measurement is performed on a two probes configuration.

- Residual thermoelectric voltage: Extraneous voltages from poor contacts or the residual thermoelectric voltages from the probe wires can contribute to the inaccuracies of the measurements. The thermoelectric voltages from the probe wires are subtracted in this setup but offset at $\Delta T=0$ can still be a source of inaccuracy. To eliminate this, $\Delta V$ vs $\Delta T$ data points are plotted and the residual voltages are removed after linear fit, and the Seebeck value is calculated from the slope. This process is repeated three times to get an average final S.

- Temperature reading corrections: Small thermoelectric voltages which may appear in any circuit and possibly in the switching network box may affect the temperature readings. Therefore, both thermocouples were calibrated to match the same temperature readings for a given temperature. This was done through a slow calibration procedure in vacuum using the furnace without any sample. It is to note that, adequate vacuum level (of the order of $10^{-5}$ Torr) is required to perform accurate and precise measurements.

- Temperature stabilization: At every stage of the Seebeck measurement, temperature stabilization is needed to be able to measure in steady state. The waiting time for temperature stabilization varies significantly with set temperatures and hence a sufficient waiting time is recommended before any measurement. Figure 8(a) shows an overall temperature profile with a step every 50 °C. For example, to get a stable global temperature (variation of temperature < 1 °C/min) at 80 °C, 100 min is necessary while at 480 °C only 10 min is enough. In a similar manner, the stabilization of the ΔT differs depending on the global temperature and requires different heater output to maintain the generated ΔT. To calibrate the required heater voltages, a set of measurements were performed as shown in Figure S6.

- Seebeck coefficient errors: The error bar on the Seebeck coefficient values come mainly from the temperature evaluation where $S = -\Delta V/\Delta T$. Because the temperature gradients are usually in the order of 2-10 °C, minimizing the error of measurement below 5 % on the temperature reading is challenging. An error of ΔT reading is estimated at 0.2 °C which would correspond to 7 % with a ΔT of 3 °C while the error is reduced to 2 % for a higher ΔT of 10 °C.
- Additional error depending on the sample: Samples showing higher variation in Seebeck values as a function of temperature can result in non-linear ΔV vs ΔT plot resulting in higher inaccuracies. For such samples, it is recommended to provide smaller ΔT to minimize the error of measurement on the voltage drop under a ΔT.

# Validation and measurement with reference samples

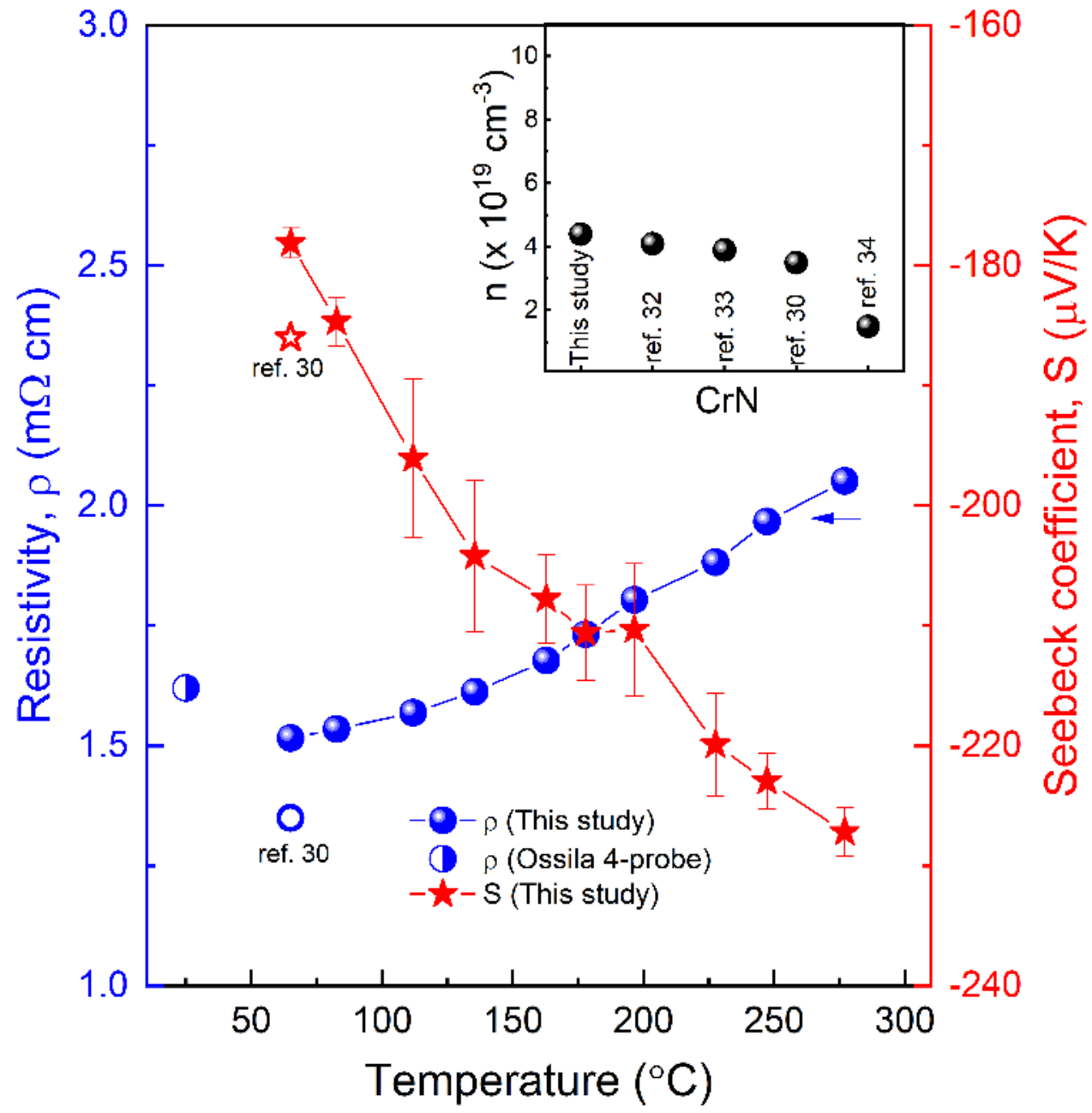

Figure 9: Temperature dependent resistivity and Seebeck coefficient for a reference CrN thin film sample along with room temperature values from literature on CrN. The inset shows room temperature values of charge carrier density in comparison with reported literature.

A temperature dependent measurement of Seebeck coefficient, resistivity and room temperature Hall effect measurement is performed on a CrN thin film (220 nm) grown on c-sapphire substrate and shown in Figure 9. The temperature dependent measurements are performed with a step size of 25 °C. The resistivity increases slightly with temperature which can be attributed to the narrow band gap of CrN and it is consistence with reported literature.[30] Additionally, to validate the results, room temperature resistivity was measured using Ossila T2001A four-point-probe station, which shows good agreement with the value measured by the setup. The absolute value of Seebeck coefficient, S increases from ~180 µV/K at room temperature to ~230 µV/K at ~275 °C. The measured resistivity and Seebeck coefficient values are compared (marked in Figure 9) with report available in literature for CrN grown under similar conditions.[30, 31] The inset of Figure 9 shows the room temperature charge carrier density, n = ~ 4 x $10^{19}$ $cm^{-3}$ which is in good agreement with reported values.[30, 32-34]

For checking the instruments performance at higher temperatures, electrical conductivity and Seebeck coefficient were measured from room temperature to 600 °C on a ~ 264 nm ScN film deposited at 750 °C on c-sapphire substrate whose growth conditions are reported elsewhere.[35] Both, electrical conductivity and Seebeck coefficient shows excellent agreement with the literature.

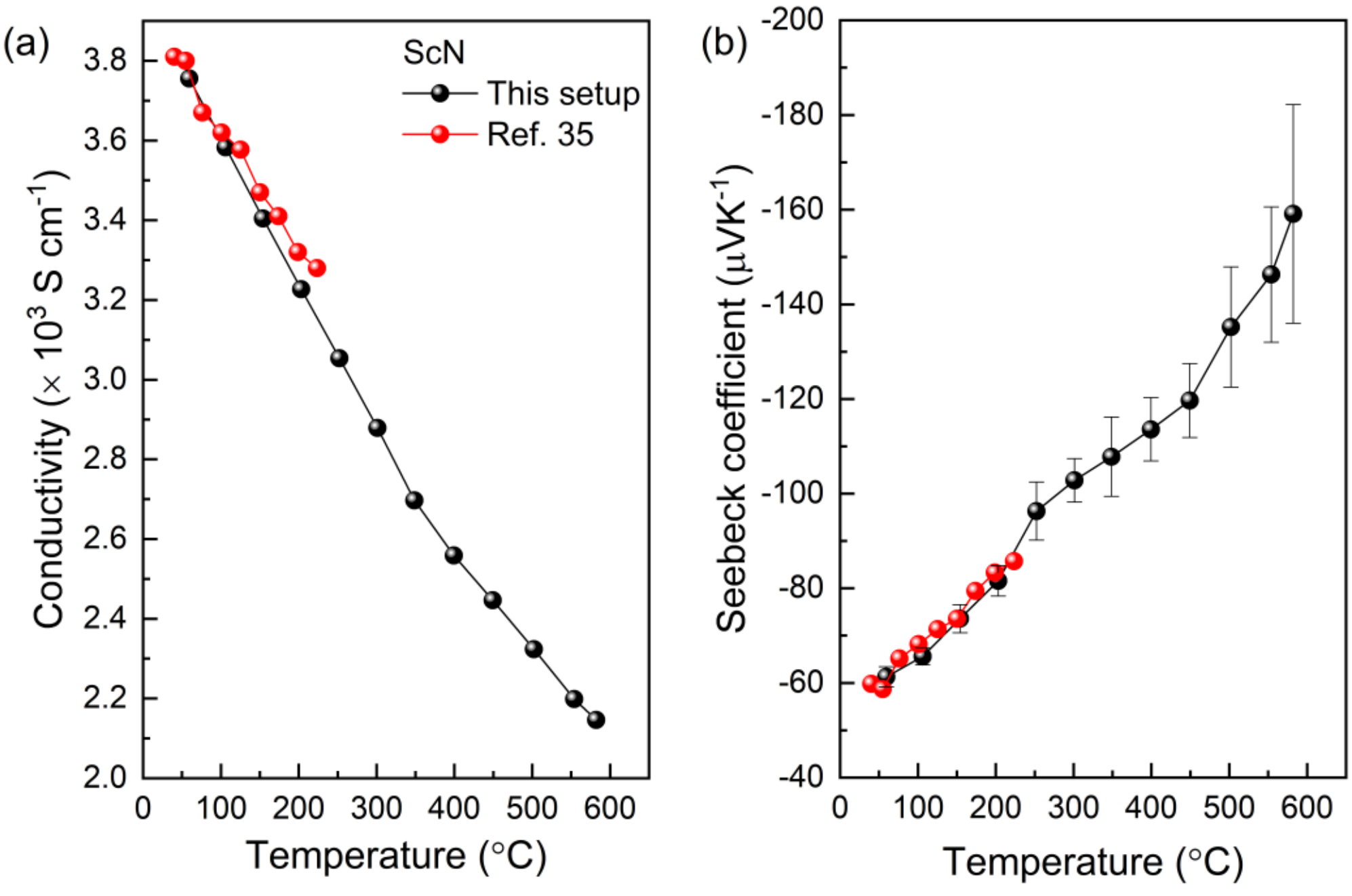

Figure 10: Measured temperature dependent (a) electrical conductivity and (b) Seebeck coefficient for a reference 264 nm ScN film grown at 750 °C on c-sapphire substrate compared with the originally reported values.[35]

Furthermore, it is important to verify the repeatability and hysteresis characteristics of electrical measurements when evaluating the performance and reliability of for any material system. Repeatable measurements ensure the stability of reproducibility of the instrumental setup, while hysteresis analysis helps identify possible measurement artifacts, contact-related effects, or intrinsic material responses that may influence the accuracy of the results. For this purpose, we used CrN film (220 nm) grown on c-sapphire substrate.

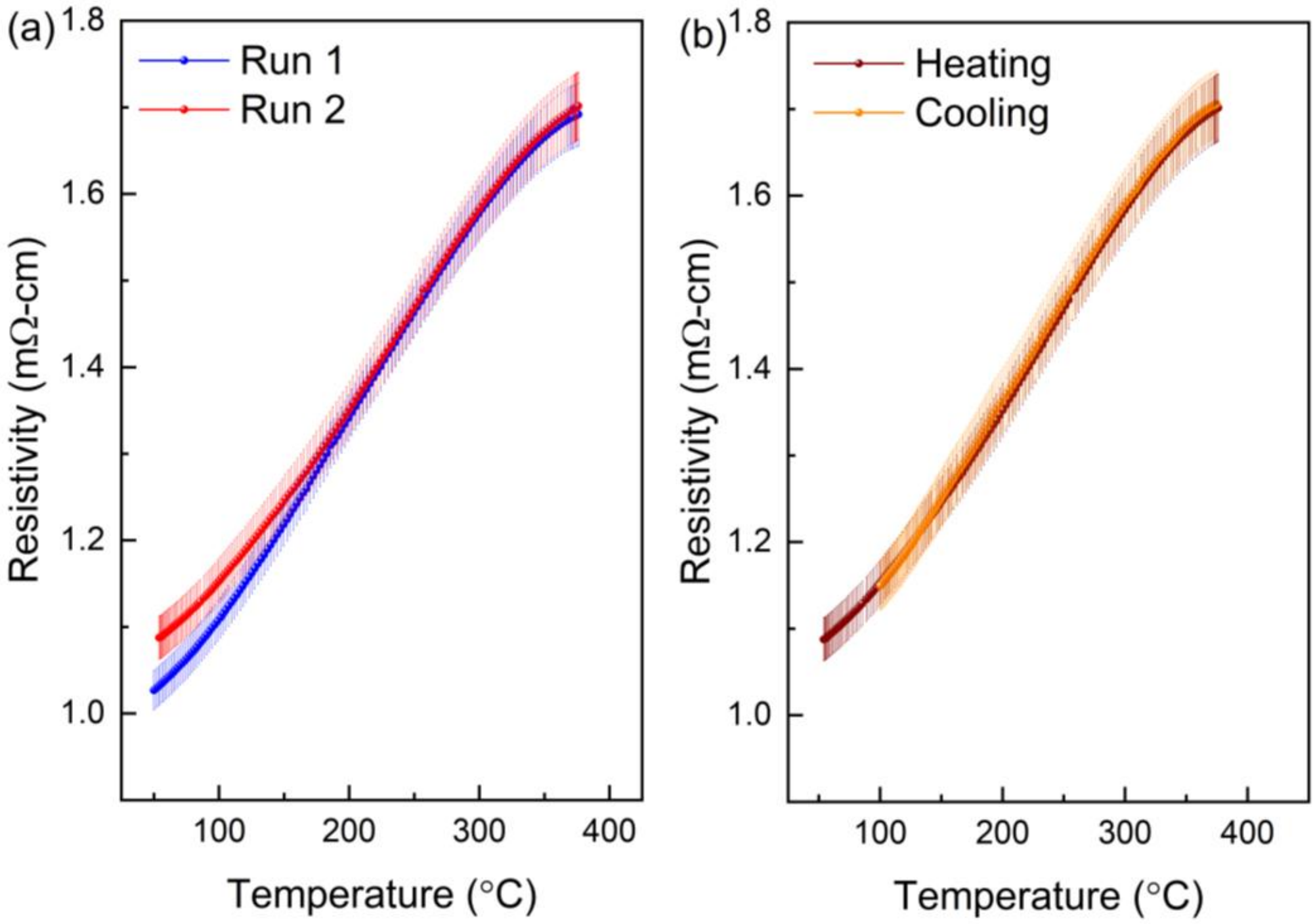

Figure 11: Measured temperature dependent (a) electrical conductivity and (b) Seebeck coefficient for a reference 264 nm ScN film grown at 750 °C on c-sapphire substrate compared with the originally reported values.

The temperature dependent resistivity measurement was repeated twice as shown in figure 10(a). The measurements show good repeatability except slight deviations around 100 °C, while no hysteresis was observed in the heating/cooling cycle (Fig. 10(b)).

## Conclusion

In conclusion, a simple modular thermoelectric measurement setup in van der Pauw configuration is developed for straightforward and simultaneous measurement of electrical resistivity and Seebeck coefficient in a wide range of temperature (25°C - 600°C) and Hall coefficients at room temperature. The setup is designed to minimize the potential errors which may arise from offset voltages, wire contributions and thermal contact resistances. Minimization of error leads to accurate measurements of voltages and temperatures which yields reliable output data. The measurement protocols are developed based on series of various optimization steps to ensure stable set temperatures and temperature gradient. The calibration of thermocouples ensures removal of residual voltages during measurements. The setup is controlled by a user friendly and fully automatic LabVIEW program. The detachable modules for Seebeck coefficient/resistivity and hall effect measurement make this setup quite versatile and provides an all-in-one (except thermal conductivity) platform for thermoelectric measurements for thin films.

## Acknowledgement

The authors acknowledge students (Evelina Waldmann, Henning Gideskog, Hilma Karlsson and Hugo Forssell) of the course Applied Physics - Bachelor Project, TFYA75 at Linköping University (2019), for their valuable input on the design and assistance on construction of the first version of the electrical setup. The authors acknowledge support from the Knut and Alice Wallenberg foundation through the Wallenberg Academy Fellows program (KAW-2020.0196) and the Swedish Research Council (VR) under Project No. 2021-03826.

## Declaration of Interest

The authors declare no competing financial interest.

## DATA AVAILABILITY

The data that support the findings of this study are available from the corresponding author upon reasonable request.

# References

[1]D. M. Rowe, *CRC Handbook of Thermoelectrics* (CRC Press).

[2]G. J. Snyder and E. S. Toberer Nature Materials **7,** (2008).

[3]J. R. Sootsman, D. Y. Chung and M. G. Kanatzidis Angewandte Chemie International Edition **48,** (2009).

[4]K. Behnia, *Fundamentals of Thermoelectricity* (Oxford University Press, 2015).

[5]J. P. Heremans and J. Martin Nature Materials **23,** (2024).

[6]J. de Boor and E. Müller Review of Scientific Instruments **84,** (2013).

[7]J. de Boor and V. Schmidt Applied Physics Letters **99,** (2011).

[8]C. Wood, A. Chmielewski and D. Zoltan Review of Scientific Instruments **59,** (1988).

[9]S. R. Sarath Kumar and S. Kasiviswanathan Review of Scientific Instruments **79,** (2008).

[10]Z. Zhou and C. Uher Review of Scientific Instruments **76,** (2005).

[11]V. Ponnambalam, S. Lindsey, N. S. Hickman and T. M. Tritt Review of Scientific Instruments **77,** (2006).

[12]B. Paul Measurement **45,** (2012).

[13]A. Guan, H. Wang, H. Jin, W. Chu, Y. Guo and G. Lu Review of Scientific Instruments **84,** (2013).

[14]F. Chen, J. C. Cooley, W. L. Hults and J. L. Smith Review of Scientific Instruments **72,** (2001).

[15]R. H. Freeman and J. Bass Review of Scientific Instruments **41,** (1970).

[16], (http://www.linseis.com for LSR-3 of LINSEIS).

[17], (http://www.ulvac.com for ZEM3 of ULVAC technologies.).

[18], (https://www.qdusa.com/ for PPMS, Quantum Design ).

[19], (http://www.mmr-tech.com for SB-100 of MMR Technologies).

[20]S. Masoumi, A. Noori and A. Pakdel Measurement **236,** (2024).

[21]J.-H. Hong, D. Kim, M.-J. Kim, S. Chung, H.-C. Shin, S.-M. Kim, K. Cho, H. S. Lee, S. Lee and B. Kang Organic Electronics **108,** (2022).

[22]R. Xiong, S. Masoumi and A. Pakdel, *Energies,* (2023).

[23]B. Dörling, O. Zapata-Arteaga and M. Campoy-Quiles Review of Scientific Instruments **91,** (2020).

[24]S. Ferreira-Teixeira, F. Carpinteiro, J. P. Araújo, J. B. Sousa and A. M. Pereira Review of Scientific Instruments **92,** (2021).

[25]O. Boffoué, A. Jacquot, A. Dauscher, B. Lenoir and M. Stölzer Review of Scientific Instruments **76,** (2005).

[26]L. J. v. d. Pauw Philips Research Reports **13,** (1958).

[27]W. K. Chan Review of Scientific Instruments **71,** (2000).

[28]J. Mackey, F. Dynys and A. Sehirlioglu Review of Scientific Instruments **85,** (2014).

[29]K. Battes, C. Day and V. Hauer Journal of Vacuum Science & Technology A **33,** (2015).

[30]H. Bouteiller, C. Poterie, R. Burcea, D. Fournier, Y. Ezzahri, S. Dubois, P. Eklund, A. le Febvrier and J.-F. Barbot Advanced Materials Interfaces**,** (2025).

[31]V. Hjort, N. K. Singh, S. Chowdhury, R. Shu, A. Le Febvrier and P. Eklund Advanced Energy and Sustainability Research **4,** (2023).

[32]C. X. Quintela, B. Rodríguez-González and F. Rivadulla Applied Physics Letters **104,** (2014).

[33]M. A. Gharavi, S. Kerdsongpanya, S. Schmidt, F. Eriksson, N. V. Nong, J. Lu, B. Balke, D. Fournier, L. Belliard, A. le Febvrier, C. Pallier and P. Eklund Journal of Physics D: Applied Physics **51,** (2018).

[34]C. X. Quintela, F. Rivadulla and J. Rivas Applied Physics Letters **94,** (2009).

[35]A. le Febvrier, S. K. Honnali, C. Poterie, T. V. Fernandes, R. Frost, V. Rogoz, M. Magnuson, F. Giovannelli, J. P. Leitão, J. F. Barbot and P. Eklund Journal of Materials Chemistry A **14,** (2026).